\RequirePackage{lineno}
\documentclass[aps,prl,twocolumn,superscriptaddress,showpacs,showkeys,amsmath,amssymb,floatfix]{revtex4}
\usepackage{booktabs,graphicx,mathrsfs,verbatim,amsmath,units}
\usepackage[colorlinks=true,citecolor=blue,filecolor=blue,linkcolor=blue,urlcolor=blue,pdftex]{hyperref}
\usepackage[usenames,dvipsnames]{color}
\usepackage{ulem}
\usepackage{color}

\newcommand\figref[1]{Fig.\,\ref{#1}}

\begin{document}	
\title{Search for Event Rate Modulation in XENON100 Electronic Recoil Data}

\newcommand{\bern}{\affiliation{Albert Einstein Center for Fundamental Physics, University of Bern, Bern, Switzerland}}
\newcommand{\bologna}{\affiliation{Department of Physics and Astrophysics, University of Bologna and INFN-Bologna, Bologna, Italy}}
\newcommand{\coimbra}{\affiliation{Department of Physics, University of Coimbra, Coimbra, Portugal}}
\newcommand{\columbia}{\affiliation{Physics Department, Columbia University, New York, NY, USA}}
\newcommand{\lngs}{\affiliation{INFN-Laboratori Nazionali del Gran Sasso and Gran Sasso Science Institute, L'Aquila, Italy}}
\newcommand{\mainz}{\affiliation{Institut f\"ur Physik \& Exzellenzcluster PRISMA, Johannes Gutenberg-Universit\"at Mainz, Mainz, Germany}}
\newcommand{\heidelberg}{\affiliation{Max-Planck-Institut f\"ur Kernphysik, Heidelberg, Germany}}
\newcommand{\munster}{\affiliation{Institut f\"ur Kernphysik, Wilhelms-Universit\"at M\"unster, M\"unster, Germany}}
\newcommand{\nikhef}{\affiliation{Nikhef and the University of Amsterdam, Science Park, Amsterdam, Netherlands}}
\newcommand{\nyuad}{\affiliation{New York University Abu Dhabi, Abu Dhabi, United Arab Emirates}}
\newcommand{\purdue}{\affiliation{Department of Physics and Astronomy, Purdue University, West Lafayette, IN, USA}}
\newcommand{\rensselaer}{\affiliation{Department of Physics, Applied Physics and Astronomy, Rensselaer Polytechnic Institute, Troy, NY, USA}}
\newcommand{\rice}{\affiliation{Department of Physics and Astronomy, Rice University, Houston, TX, USA}}
\newcommand{\shanghai}{\affiliation{Department of Physics \& Astronomy, Shanghai Jiao Tong University, Shanghai, China}}
\newcommand{\subatech}{\affiliation{SUBATECH, Ecole des Mines de Nantes, CNRS/In2p3, Universit\'e de Nantes, Nantes, France}}
\newcommand{\torino}{\affiliation{INFN-Torino and Osservatorio Astrofisico di Torino, Torino, Italy}}
\newcommand{\ucla}{\affiliation{Physics \& Astronomy Department, University of California, Los Angeles, CA, USA}}
\newcommand{\weizmann}{\affiliation{Department of Particle Physics and Astrophysics, Weizmann Institute of Science, Rehovot, Israel}}
\newcommand{\zurich}{\affiliation{Physik-Institut, University of Zurich, Zurich, Switzerland}}

\author{E.~Aprile}\columbia
\author{J.~Aalbers}\nikhef
\author{F.~Agostini}\lngs\bologna
\author{M.~Alfonsi}\mainz\nikhef
\author{M.~Anthony}\columbia
\author{L.~Arazi}\weizmann
\author{K.~Arisaka}\ucla
\author{F.~Arneodo}\nyuad
\author{C.~Balan}\coimbra
\author{P.~Barrow}\zurich
% ...

\collaboration{The XENON Collaboration}\noaffiliation

\date{\today}

\begin{abstract} 
We have searched for periodic variations of the electronic recoil event rate in the $(2-6)$\,keV energy range recorded between February 2011 and March 2012 with the XENON100 detector, adding up to 224.6 live days in total. 
Following a detailed study to establish the stability of the detector and its background contributions during this run, we performed an un-binned profile likelihood analysis to identify any periodicity up to 500 days. 
We find a global significance of less than $1\,\sigma$ for all periods suggesting no statistically significant modulation in the data. While the local significance for an annual modulation is $2.8\,\sigma$, the analysis of a multiple-scatter control sample and the phase of the modulation disfavor a dark matter interpretation. 
The DAMA/LIBRA annual modulation interpreted as a dark matter signature with axial-vector coupling of WIMPs to electrons is excluded at $4.8\,\sigma$.

\end{abstract}
\pacs{
 95.35.+d, 
 14.80.Ly, 
 29.40.-n, 
}
\keywords{Dark Matter, WIMPs, Annual Modulation, Direct Detection, Liquid Xenon}
\maketitle

The XENON100 experiment~\cite{Aprile:2012instr} is designed to search for dark matter in the form of Weakly Interacting  Massive Particles (WIMPs)~\cite{Bertone:2005theory} by detecting WIMP-induced nuclear recoils (NRs) with a liquid xenon (LXe) time projection chamber. The resulting event rate in any dark matter detector is expected to be annually modulated due to the relative motion between the Earth and the dark matter halo of the Milky Way~\cite{Freese:2012modulation}.
The modulation of the low energy (low-E), ($2-6$)\,keV, event rate in the DAMA/LIBRA experiment~\cite{Bernabei:dama} is currently the only long-standing claim for a positive dark matter detection. 
Under typical astrophysical and particle physics assumptions, this claim is however challenged by the non-observation of WIMP-induced NRs of several other experiments using different target materials and detector technologies~\citep[e.g.][]{Aprile:2012run10,Agnese:cdmslowe,Akerib:lux}, most with considerably lower radioactive backgrounds.

An alternative explanation is that the DAMA/LIBRA annual modulation is due to electronic recoils (ERs) from WIMPs which have axial-vector couplings to electrons~\cite{Kopp:leptonic,Aprile:dcpaper}. The stable performance of XENON100 over a period of more than one year offers the opportunity to test this hypothesis with a different detector operated for the first time in the same underground site, the Laboratori Nazionali del Gran Sasso (LNGS), Italy.

For this analysis we use the 224.6~live days of XENON100 dark matter data accumulated from February 28, 2011 to March 31, 2012, previously used to search for  spin-independent~\cite{Aprile:2012run10} and spin-dependent~\cite{Aprile:2013sd} WIMP-induced NRs as well as for axion-induced ERs~\cite{Aprile:axion} and a comparison with DAMA/LIBRA using the average ER rate~\cite{Aprile:dcpaper}.

The ER energy and uncertainty therein is inferred from the prompt scintillation light signal (S1), as in~\cite{Aprile:axion}, using the NEST model (v0.98)~\cite{nest} fit to independent light yield calibration measurements~\cite{Aprile:erquench,laura:erquench}. The overall uncertainty on the ER energy scale is dominated by the spread of the low energy measurements in~\cite{Aprile:erquench,laura:erquench} and is estimated to be 14\% at 2\,keV and 9\% at 6\,keV.

We use the same S1 range of ($3-30$)\,photoelectrons (PE) as in~\cite{Aprile:2012run10,Aprile:xe100analysis}, but divided into two ranges. The low-E range ($3-14$)\,PE corresponds to ($2.0-5.8$)\,keV and thus
covers the energy interval where the DAMA/LIBRA experiment observes a modulation signal. The higher energy range, ($14-30$)\,PE, corresponds to ($5.8-10.4$)\,keV and is used as a sideband control sample. 

\begin{figure}
\centering
\includegraphics[width=1\columnwidth]{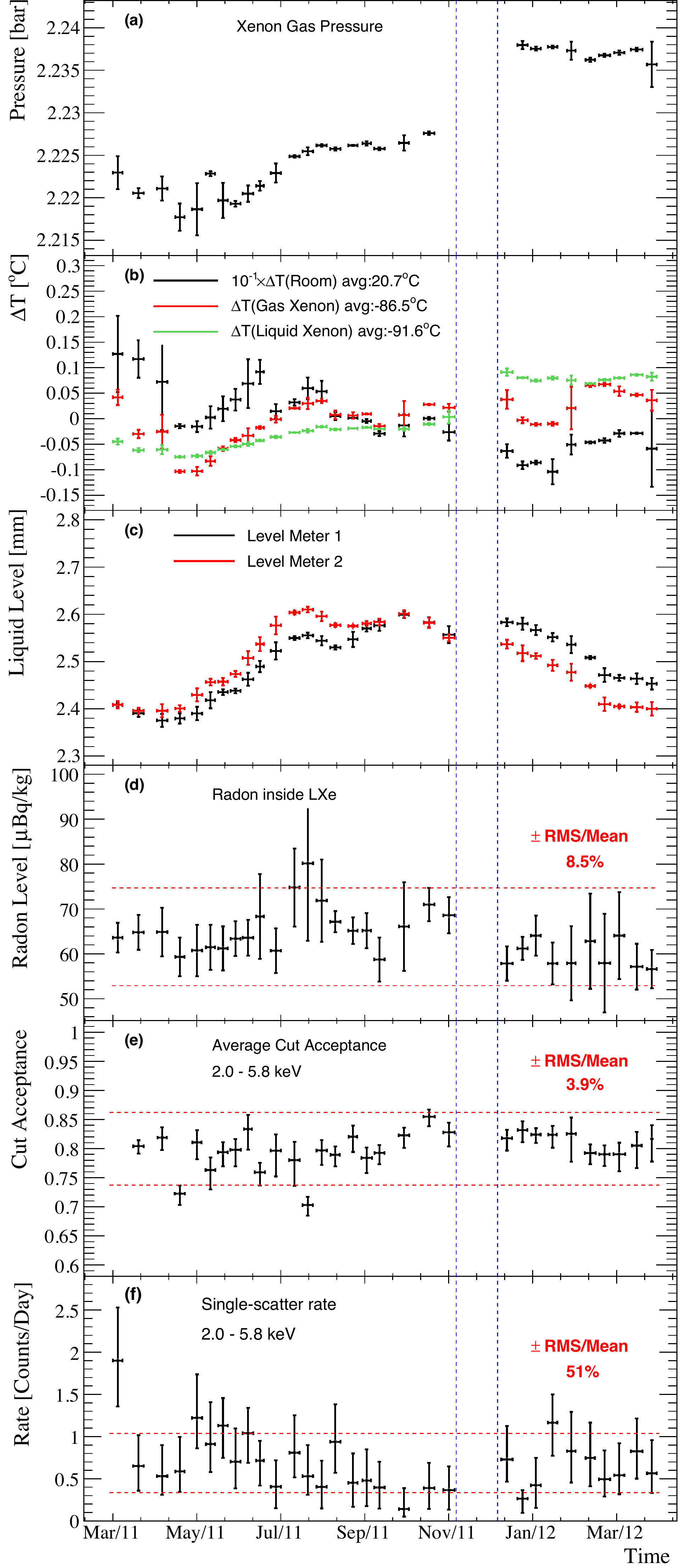}
\caption{\label{stability} 
Temporal evolution of the relevant XENON100 detector parameters studied for this analysis.
The dashed blue lines indicate a detector maintenance period.
(a-c)  Xe pressure, LXe temperature and LXe level. (d) Radon level in the 34\,kg LXe fiducial mass, as measured via {\it in situ} alpha spectroscopy. (e) Average cut acceptance in the low-E range of ($2.0-5.8$)\,keV, as derived from weekly ER calibrations. (f) ER event rate in the 34\,kg fiducial mass for single-scatters in the low-E range.\vspace{-0.12in}
}
\end{figure}

Low-E single-scatter events in the 34~kg fiducial mass, as expected from dark matter interactions, are selected using the same cuts as in~\cite{Aprile:2012run10}. While these cuts were defined to select valid NR events, they also have high efficiency for ERs~\cite{Aprile:axion}, and result in 153 events distributed in time as shown in \figref{stability}\,(f). The cut acceptances in the energy ranges considered here have been derived following the procedure in~\cite{Aprile:xe100analysis} using ER calibration data ($^{60}$Co and $^{232}$Th) taken on a weekly basis. The time variation of the acceptance, shown in \figref{stability}\,(e), is incorporated in the analysis by linearly interpolating between the data points. We have verified that our conclusions remain unaffected when adopting different methods of cut acceptance interpolation in time.

The design of XENON100 incorporates many sensors of various types to monitor the long-term stability of the detector. A total of 15 parameters were investigated, of which a subset with the highest potential impact on detector signals is shown in \figref{stability}\,(a--d). The absolute pressure of the gas above the LXe has a mean value of 2.23\,bar with a maximum variation of 0.02\,bar over the entire period (\figref{stability}\,(a)). The temperature sensors located at various positions within the detector exhibit a mean value varying from $-86.5\,^{\circ}{\rm C}$ (in the Xe gas) to $-91.6\,^{\circ}{\rm C}$ (bottom of the LXe) with a maximum variation of less than $0.17\,^{\circ}{\rm C}$ for each sensor. The ambient temperature in the XENON100 room has a mean value of $20.7\,^{\circ}{\rm C}$ with a maximum variation of $3.7\,^{\circ}{\rm C}$ (\figref{stability}\,(b)). The LXe level, monitored by two capacitive sensors, shows a maximum variation of 0.22\,mm during the entire period (\figref{stability}\,(c)).

To identify potential correlations between detector parameters and ER rate, we calculate the linear (Pearson) and non-linear (Spearman-Rank) correlation coefficients for the two energy ranges studied, and for both single-scatter and multiple-scatter events. The latter are defined as events with a single-scatter in the fiducial region plus an additional S1 coincident signal in the LXe veto.
The 99\,kg LXe veto has an energy threshold of $\sim$100\,keV, thus multiple-scatter events are dominated by high-energy scatters from $\gamma$ rays~\cite{Aprile:2012instr,Aprile:ermc}.
Of all the parameters studied, two were found to give a non-correlation $p$-value smaller than 0.001.
The first parameter is the LXe level, which shows a negative linear and non-linear correlation with the low-E single-scatter rate. The second parameter is the Xe gas temperature, which shows a negative linear correlation with the low-E multiple-scatter rate. 
As expected, the LXe level and gas temperature were also found to be correlated with each other and with the room temperature. 
A change in the LXe level, gas pressure and temperature can potentially affect the observed size and width of the secondary scintillation signal, S2, which is a measure of the ionization electrons liberated in the interaction. The overall observed variation of the S2 signal is less than 5\%~\cite{Aprile:singleelectron}, while the majority of events have S2~$>$~1000 PE, much larger than the trigger threshold of 150~PE.
Consequently, a detailed inspection of the S2-dependent cuts shows that their performance is unaffected. Hence the correlation with event rate is possibly a coincidence and, regardless, does not impact our statistical analysis for periodicity described below.

The impact of decaying radioactive isotopes on the low-E ER rate is also considered for this analysis. These sources can be subdivided into external sources of $\gamma$-radiation from peripheral materials and $\beta$-radiation from the decay of radioactive Rn and $^{85}$Kr distributed in the LXe volume.

Of the relevant external $\gamma$-sources in the detector and shield materials, only $^{60}$Co ($t_{1/2}=5.27$\,y) decays on a timescale sufficiently short to potentially cause an observable change in the event rate during the time period of this study. However, the decrease in activity is found to reduce the single-scatter low-E ER rate by less than 1$\%$ of its average value, based on a Monte Carlo (MC) simulation using the measured activity level from \cite{Aprile:ermc}. Hence we assume the external $\gamma$-background to be constant for this analysis.

The short-lived isotopes $^{222}$Rn and $^{220}$Rn are constantly produced as part of the primordial $^{238}$U/$^{232}$Th decay chains and are present in the air of the room and shield cavity, as well as inside the LXe due to emanation from inner surfaces. Radon decays outside the detector, measured by commercial Rn monitors in the room, contribute negligibly to the event rate in the fiducial mass since the emitted radiation is absorbed by the shield and outer detector materials. The concentration of Rn and subsequent decay products dispersed in the LXe is continuously monitored via examination of both $\alpha$-decays and $\beta$-$\gamma$ delayed coincidence events~\cite{weberthesis}. This analysis shows that $^{222}$Rn from the $^{238}$U chain is uniformly distributed in the volume while $^{220}$Rn from the $^{232}$Th chain is negligible. 
The time-variation of the $^{222}$Rn level is shown in \figref{stability}\,(d) and exhibits a specific activity of ($63\pm1$)\,$\mu$Bq/kg. This level corresponds to a low-E ER contribution of 
($1.11\pm0.02$)~events/(keV $\cdot$ tonne $\cdot$ day) as determined by MC simulation~\cite{Aprile:ermc}. 
The 8.5$\%$ fluctuation of the $^{222}$Rn level corresponds to a less than 2\% variation of the average rate and is thus negligible compared to the observed rate fluctuation of 51\% shown in \figref{stability}\,(f). In addition, no time correlation is found by calculating the linear and non-linear correlation coefficients between the low-E ER rate and the Rn level. Therefore the evolution of the $^{222}$Rn level in time is not included in the statistical analysis below.

The other internal contamination, $^{85}$Kr, is also present in air. The concentration of $^{\mathrm{nat}}$Kr in the LXe during the period studied here was determined on November 17, 2011 to be $(14\pm2)$~parts per trillion using the rare gas mass spectrometer (RGMS) method~\cite{Aprile:2012run10,rgms}. However, it became evident after the end of the run that a small air leak in the Xe gas purification system had allowed Rn and Kr atoms to diffuse into the LXe. The leakage rate into the sensitive volume was estimated from a study of the time correlation between the external and internal concentrations of $^{222}$Rn~\cite{weberthesis}, including three RGMS measurements of $^{\mathrm{nat}}$Kr spread over the course of several months during the following run. 
Assuming a constant $^{\mathrm{nat}}$Kr concentration in air, the linear increase in time of $^{\mathrm{nat}}$Kr in the LXe was found to be proportional to the integrated number of additional $^{222}$Rn decays due to the air leak. The linear increase of the single-scatter ER rate from $^{85}$Kr has a slope $K = (2.54\pm0.53)\times10^{-3}$~events/(keV $\cdot$ tonne $\cdot$ day)/day 
assuming a $^{85}$Kr/$^{\mathrm{nat}}$Kr ratio of $2 \times 10^{-11}$~\cite{rgms}.
This time-dependent background results in an expected total increase of $(0.10 \pm 0.02)$ events/day at low-E over the course of one year, which is taken into account in the following statistical analysis.

To determine the statistical significance of a periodic time dependence in the event rate, we implement an un-binned profile likelihood (PL) method~\cite{pltest}, which incorporates knowledge of the time variation of detector parameters and radioactive backgrounds as described above. The event rate for a given energy range is described by
\begin{linenomath}
\begin{equation}
\centering
f(t)=\epsilon(t)\left(C+Kt+A\cos\left(2\pi \frac{(t-\phi)}{P}\right)\right)\textnormal{,}
\label{hpdf}
\end{equation}
\end{linenomath}
where $\epsilon$ is the corresponding average cut acceptance, interpolated from the measurements described above, $C$ is the constant component of the event rate, $Kt$ is the linearly increasing contribution from $^{85}$Kr, and $A$ is the modulation amplitude with period $P$ and phase $\phi$. Eq.~(\ref{hpdf}) is then normalized to take into account the time distribution of the dark matter data used for the analysis here, and thus becomes the probability density, $\tilde{f}(t)$, of observing an event occurring at time $t$, in days relative to January 1, 2011. The null hypothesis, no periodicity, is given by Eq.~(\ref{hpdf}) with $A = 0$.

The likelihood function used in the PL method is
\begin{linenomath}
\begin{equation}
\begin{aligned}
\mathcal{L} =\left(\prod_{i=1}^{n}\tilde{f}(t_{i})\right) \textnormal{Poiss}\left(n|N_{\rm{exp}}(E)\right)
 {\mathcal{L}_{\epsilon}} {\mathcal{L}_{K}} {\mathcal{L}_{E}}\textnormal{,}
\end{aligned}
\label{likelihood}
\end{equation}
\end{linenomath}
where $n$ and $N_{\rm{exp}}(E)$ are the total number of observed and expected events and $E$ is the energy in keV. Nuisance parameters corresponding to the uncertainties in $\epsilon$, $K$, and $E$ are constrained by the Gaussian penalty terms, $\mathcal{L}_{\epsilon}$, $\mathcal{L}_{K}$, and $\mathcal{L}_{E}$, respectively.  These penalty terms have widths $\sigma_\epsilon$ defined by the statistical errors of the acceptance as determined by weekly calibration measurements, $\sigma_K = 0.53 \times 10^{-3}$ events/(keV $\cdot$ tonne $\cdot$ day)/day, and $\sigma_E$ taken from Fig.\,2 of \cite{Aprile:axion}, respectively. The maximum profiled likelihoods are denoted by $\mathcal{L}_{0}(C_{0})$ for the null hypothesis and by $\mathcal{L}_{1}(C_{1},A,\phi)$ for the periodic hypothesis.

\begin{figure}
\centering
\includegraphics[width=1\columnwidth]{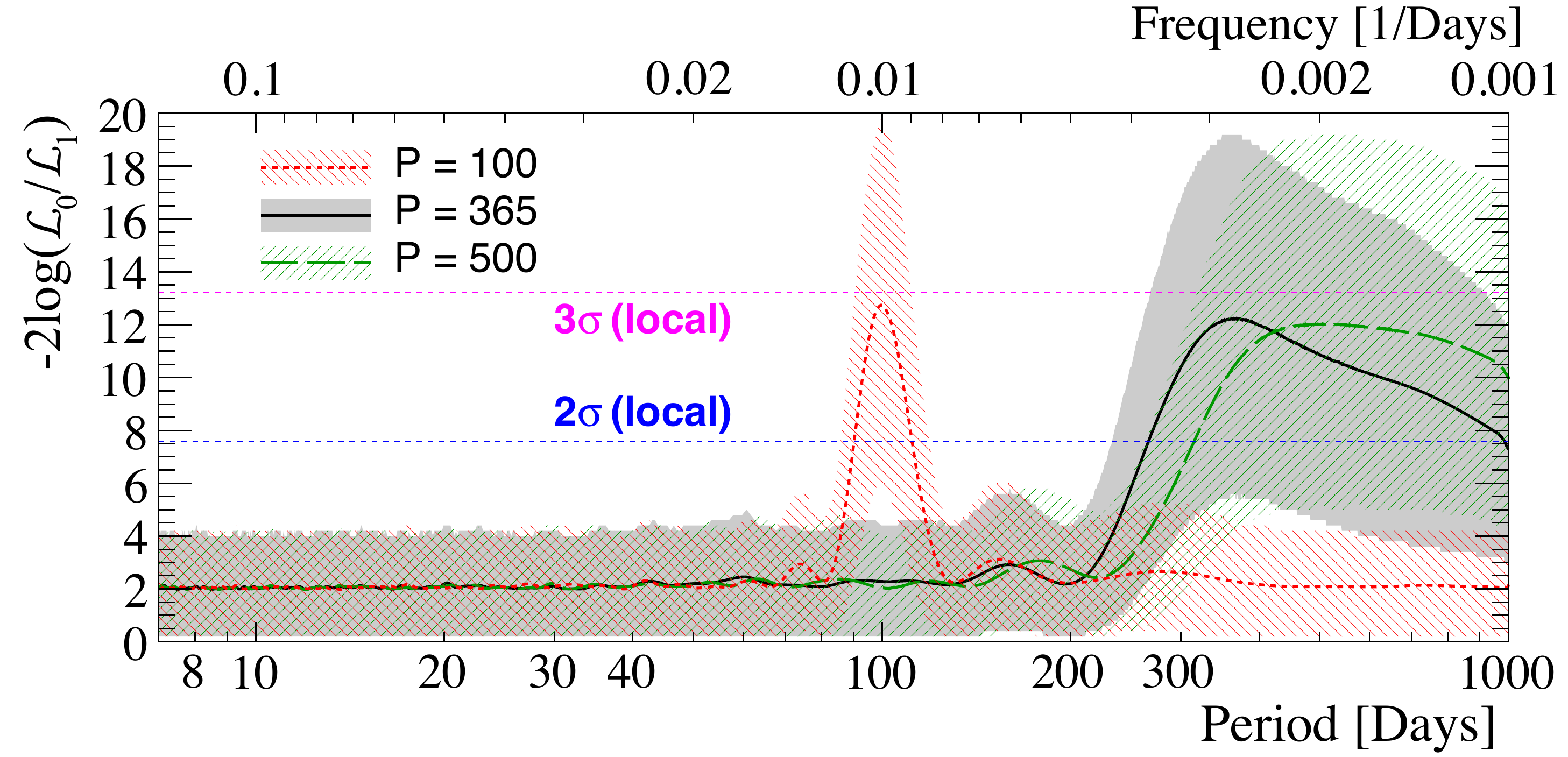}
\vspace{-0.25in}
\caption{\label{plres} The expected mean (solid lines) and central 68.3\% region (shaded bands) of $-2\log(\mathcal{L}_{0}/\mathcal{L}_{1})$ as a function of period for simulated data with a fixed average rate $C=6.0$~events/(keV $\cdot$ tonne $\cdot$ day),
linear increase in rate $K = (2.54\pm0.53)\times10^{-3}$~events/(keV $\cdot$ tonne $\cdot$ day)/day,
amplitude $A=2.7$~events/(keV $\cdot$ tonne $\cdot$ day),
and three periods $P$ [days]. Uncertainties on all parameters are taken into account. The horizontal local significance lines are derived from the null hypothesis tests described in the text and shown here for comparison to \figref{unbinnedpl}.\vspace{-0.12in}
}
\end{figure}

The significance of a particular period, for example one year, is referred to as the local significance. The corresponding test statistic is the log-likelihood ratio, $-2\log(\mathcal{L}_{0}/\mathcal{L}_{1})$, which quantifies the incompatibility between the null and periodic hypotheses. MC simulations show that this test statistic is well-described by a $\chi^{2}$-distribution with two degrees of freedom. When searching for a modulation signal across a range of periods, the global significance, that is the maximum of the local test statistics in the range, should be referenced. The local and global significances quoted are both one-sided.

\begin{figure}
\centering
\includegraphics[width=1\columnwidth]{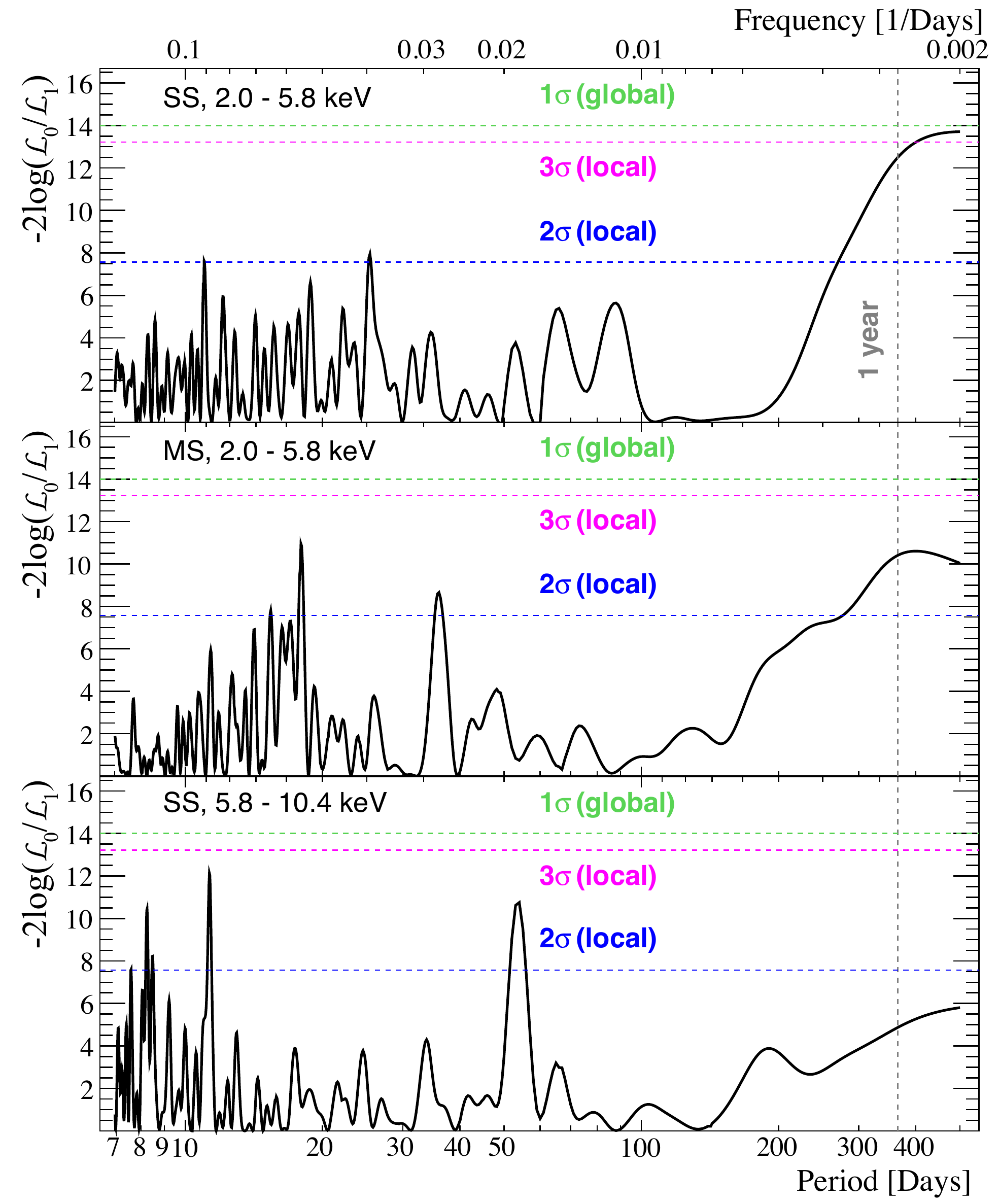}
\vspace{-0.25in}
\caption{\label{unbinnedpl} $-2\log(\mathcal{L}_{0}/\mathcal{L}_{1})$ as a function of modulation period for single-scatters (SS) in the low-E region (top), multiple-scatters (MS) in the low-E region (middle) and single-scatters (SS) in the higher energy region (bottom). The phase is unconstrained.
\vspace{-0.12in}
}
\end{figure}

Simulated data were used to assess the discovery potential of the PL analysis to periodic components in the single-scatter data at low-E.
Several sets of 153 simulated events were generated by drawing from the same live-time distribution as the actual data while varying the nuisance parameters according to their constraints in Eq.~\ref{likelihood}, and assuming the periodic hypothesis with a fixed period, amplitude and average rate. 
The expected significance is shown in \figref{plres} for three periods with an amplitude of $2.7$~events/(keV $\cdot$ tonne $\cdot$ day) and average rate of 6.0~events/(keV $\cdot$ tonne $\cdot$ day), selected to facilitate comparison with the best-fit results below. The minimum period considered is 7\,days, since the cut acceptance is derived from weekly calibration measurements. The resolution on the reconstructed period becomes worse with increasing period, evident from the broadening of the peaks and a characteristic plateau for periods $\gtrsim 500$\,days. Hence the study of the data in \figref{unbinnedpl} was limited to periods between 7 and 500\,days. Adding the previous 100.9\, live days of data~\cite{Aprile:2011run08} to this analysis does not considerably increase the significance of the study due to its higher background rate from $^{85}$Kr and the uncertainty therein.

In addition to the un-binned PL analysis, a $\chi^2$-test following~\cite{chi2testqian} and a Lomb-Scargle (LS) periodogram~\cite{lstest} were carried out using binned data. For both tests, a strong binning dependence of the result is observed. This dependence, as well as the unavoidable information loss when using any bin-dependent method, limits the power of these tests compared to the un-binned PL analysis. This fact must be taken into account when using the data in \figref{stability}\,(f) for further analysis. Nevertheless, the local and global significances are in agreement with the results of the PL analysis and the tests provide a consistency check. 

WIMP interactions in the LXe are expected to produce single-scatter events. The PL spectrum of the single-scatter data covering the DAMA/LIBRA energy region ($2.0-5.8$\,keV) is shown in the top panel of \figref{unbinnedpl}. A rise in significance is observed at long periods with a local significance of $2.8\,\sigma$ at one year and a global significance below $1\,\sigma$ for all periods. MC simulations with $P=100$\,days in \figref{plres} show that the rise of significance at large periods in the measured data is not an artifact of the statistical method.

Low-E multiple-scatter events are used as a background-only control sample. The PL spectrum (middle panel of \figref{unbinnedpl}) shows a rise in significance at long periods, similar to that for single-scatters, with a local significance of $2.5\,\sigma$ at one year and a global significance below $1\,\sigma$ at all periods.

As WIMPs are expected to produce signals primarily at low-E, the higher energy range ($5.8-10.4$\,keV) is used as a sideband control sample. In addition, DAMA/LIBRA did not observe a modulation above 6\,keV. The PL spectrum (bottom panel of \figref{unbinnedpl}) shows no prominent rise in significance at long periods, in contrast to that seen at low-E, and the local significance is $1.4\,\sigma$ at one year.  

\begin{figure}
\centering
\vspace{0.02in}
\includegraphics[width=1\columnwidth]{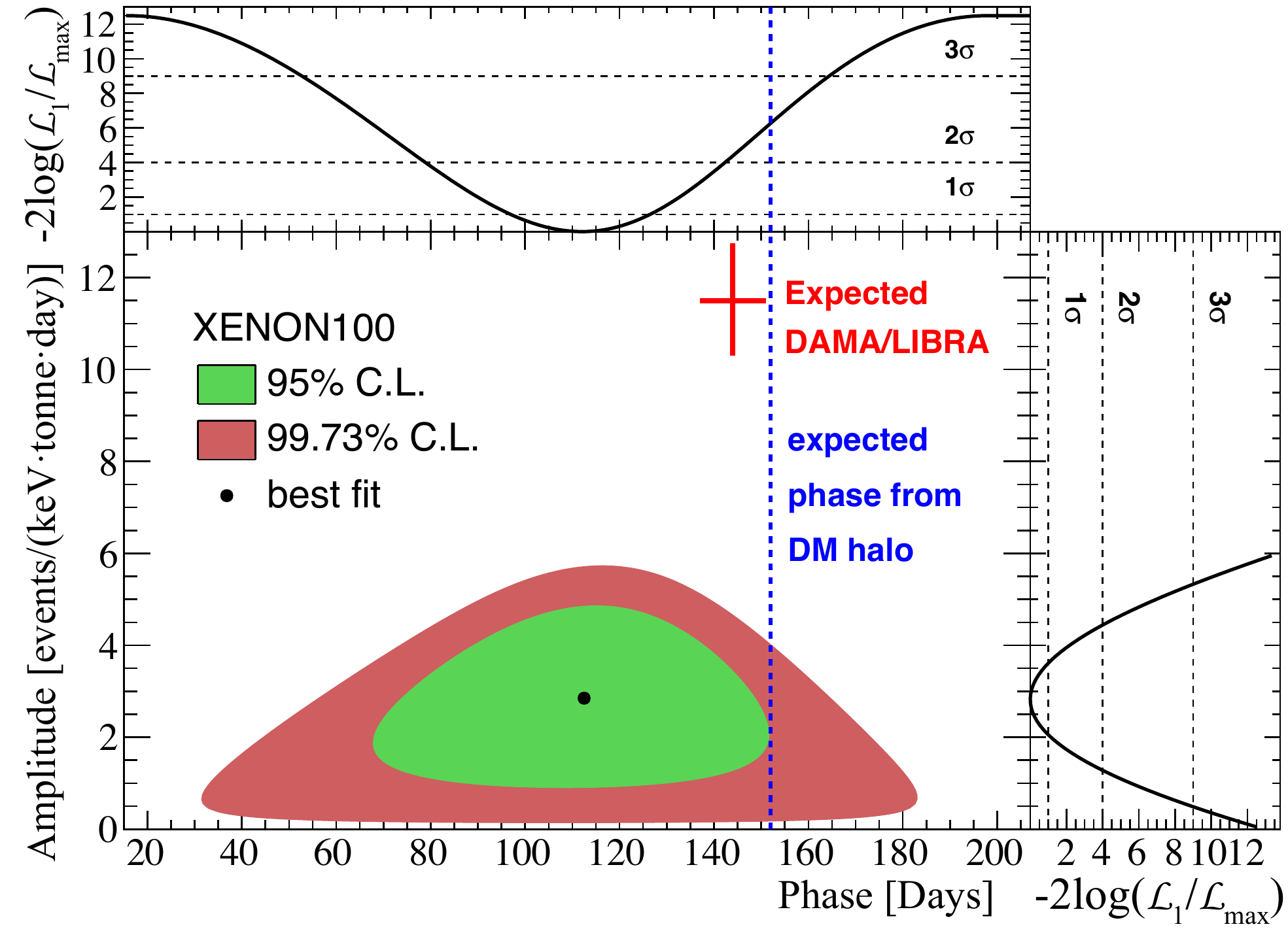}
\vspace{-0.3in} 
\caption{\label{plphase} The XENON100 best-fit, $95\%$ and $99.73\%$ confidence level contours as a function of amplitude and phase relative to January 1, 2011 for period $P~=~1$~year. The expected DAMA/LIBRA signal with statistical uncertainties only and the phase expected from a standard dark matter (DM) halo are overlaid for comparison. Top and side panels show $-2\log(\mathcal{L}_{1}/\mathcal{L}_{max})$ as a function of phase and amplitude, respectively, along with two-sided significance levels.\vspace{-0.12in}  
}
\end{figure}

The best-fit parameters and uncertainties are determined from PL scans. For an assumed annual modulation signal (fixing $P=365.25$\,days) in the low-E single scatter data, we obtain  
$C_1 = (5.5\pm0.6)$~events/(keV $\cdot$ tonne $\cdot$ day)
(for reference, 
$C_0 = 6.0$~events/(keV $\cdot$ tonne $\cdot$ day)), 
$A = (2.7\pm0.8)$~events/(keV $\cdot$ tonne $\cdot$ day),
and $\phi = (112\pm15)$~days, peaked at April\,22. \figref{plphase} shows the corresponding confidence level contours as a function of modulation amplitude and phase. The simulations in \figref{plres} show that the rise in significance at long periods in the low-E single- and multiple-scatter data could be explained by a modulating component with a period $\gtrsim$300 days.
However, the best-fit phase disagrees with the expected phase from a standard dark matter halo (152~days) at a level of $2.5\,\sigma$ based on the 1D PL scan as shown in top panel of \figref{plphase}. Furthermore, the rise in significance at long periods is evident in both single- and multiple-scatter data, also disfavoring a WIMP interpretation. Allowing the parameter $K$ to float freely to unphysical negative values, given the measured $^{85}$Kr level, decreases the significance of large periods and strengthens the exclusion limit discussed below.

The XENON100 data can constrain the dark matter interpretation of the annual modulation observed by DAMA/LIBRA, as shown in \figref{plphase}, for certain models producing ERs. Such constraints were previously imposed using the average ER event rate in XENON100~\cite{Aprile:dcpaper}. Here we use the full time-dependent rate information to directly compare with the expected DAMA/LIBRA annual modulation signal in our detector. The expected S1 spectrum in XENON100 is derived from the DAMA/LIBRA residual modulation spectrum (Fig.\,8 in \cite{Bernabei:dama}) following the approach described in~\cite{Aprile:dcpaper}, assuming the signals are from WIMP-electron scattering through axial-vector coupling~\cite{Kopp:leptonic, Aprile:dcpaper}. The expected annual modulation amplitude in the low-E range in XENON100 is then calculated as $(11.5 \pm 1.2(\rm{stat}) \pm 0.7(\rm{syst})) $~events/(keV $\cdot$ tonne $\cdot$ day), with statistical uncertainty from the reported DAMA/LIBRA spectrum and systematic uncertainty from the energy conversion in XENON100. To compare this expected signal with our data, the phase $\phi$ in Eq.~(\ref{hpdf}) is set to ($144\pm7$)\,days~\cite{Bernabei:dama}, constrained by an additional Gaussian term, $\mathcal{L_{\phi}}$, in Eq.~\ref{likelihood}. The resulting PL analysis of our data disfavors the expected DAMA/LIBRA annual modulation at $4.8\,\sigma$.

In summary, XENON100 has demonstrated for the first time that LXe dual-phase time projection chambers can be operated with sufficient long-term stability to enable searches for periodic signals for periods up to and greater than one year. The detector parameters investigated were found to be very stable, and most show no correlation with the measured low-E ($2.0-5.8$\,keV) single-scatter ER event rate. Although the LXe level and Xe gas temperature show a correlation with this rate, no evidence was found of a direct impact on the cut performance. A time varying cut acceptance and background from $^{85}$Kr are included in the search for event rate modulation. In the 224.6~live days of XENON100 data taken over the course of more than one year, a rising significance at long periods is observed for low-E single- and multiple-scatter events with the most likely period being $\gtrsim$450 days. An explicit search for annual modulation in the ER rate gives a $2.8\,\sigma$ local significance with a maximum rate on April~$22 \pm 15$ days. This phase disfavors an annual modulation interpretation due to the standard dark matter halo at $2.5\,\sigma$. Furthermore, our results disfavor the interpretation of the DAMA/LIBRA annual modulation signal as being due to WIMP-electron scattering through axial-vector coupling at $4.8\,\sigma$.

We gratefully acknowledge support from NSF, DOE, SNF, FCT, Region des Pays de la Loire, STCSM, NSFC, DFG, MPG, Stichting FOM, the Weizmann Institute of Science, EMG, ITN Invisibles, and INFN. We are grateful to LNGS for hosting and supporting the XENON project.

\end{document}